\documentclass{PoS}

\usepackage{epsfig,multicol,multirow}
\usepackage{amsmath,subfigure,latexsym,amssymb}

\def\lsi{\lesssim}
\def\gsi{\gtrsim}

\newcommand{\beq}{\begin{equation}}
\newcommand{\eeq}{\end{equation}}

\def\lsi{\raise0.3ex\hbox{$<$\kern-0.75em\raise-1.1ex\hbox{$\sim$}}}
\def\gsi{\raise0.3ex\hbox{$>$\kern-0.75em\raise-1.1ex\hbox{$\sim$}}}

\def\backder{\raise1.4ex\hbox{$\leftarrow$\kern-0.75em\raise-1.4ex\hbox{$\partial$}}}

\def\forder{\raise1.4ex\hbox{$\rightarrow$\kern-0.75em\raise-1.4ex\hbox{$\partial$}}}

\newcommand{\backderi}{\mathop{\backder}}
\newcommand{\forderi}{\mathop{\forder}}

\newcommand{\NN}{{\kern+.25em\sf{N}\kern-.78em\sf{I} \kern+.78em\kern-.25em}}

\PoS{PoS(LAT2005)264}

\title{The Scaling of QED in a Non-Commutative Space-Time}

\ShortTitle{Scaling of non-commutative QED}

\author{\speaker{Jan Volkholz }$^{~a}$,
  Wolfgang Bietenholz$^{~a}$, Jun Nishimura$^{~b,c}$ and 
  Yoshiaki Susaki$^{~b,d} ~~ $\thanks{{Preprint ~ HU-EP-05/44}} \\
  \ \\
  $^{a}$ Institut f\"{u}r Physik \\ 
  ~~ Humboldt Universit\"{a}t zu Berlin \\
  ~~ Newtonstr.\ 15, D-12489 Berlin, Germany \\
  \ \\
  $^{b}$ High Energy Accelerator Research Organization (KEK) \\
  ~~ 1-1 Oho, Tsukuba 305-0801, Japan \\
  \ \\
  $^{c}$ Department of Particle and Nuclear Physics \\
  ~~ Graduate University for Advanced Studies (SOKENDAI) \\
  ~~ 1-1 Oho, Tsukuba 305-0801, Japan \\
  \ \\
  $^{d}$ Institute of Physics, University of Tsukuba, \\
  ~~ Tsukuba, Ibaraki 305-8571, Japan \\

  E-mail: \email{volkholz@physik.hu-berlin.de,
  bietenho@physik.hu-berlin.de, \\ 
  jnishi@post.kek.jp, susaki@post.kek.jp}}

\abstract{We present results of numerical simulations for pure $U(1)$ gauge
theory in a non-commutative space. The theory is mapped onto a dimensionally
reduced matrix model, which renders its numerical treatment feasible.
New data on large lattices reveal the scaling of Wilson loops and their
correlation functions
in the simultaneous limit to the continuum and to infinite volume,
at fixed non-commutativity. In this on-going project we are
particularly interested in the IR behaviour, the ``photo-ball''
spectrum and in the photon dispersion relation.}

\FullConference{XXIIIrd International Symposium on Lattice Field Theory\\

                 25-30 July 2005\\

                 Trinity College, Dublin, Ireland}

\begin{document}

\section{Non-commutative $U(1)$ gauge theory}\label{section1}


Somewhat overshadowed by the celebrations for Einstein's works
of 1905, we are also cele\-brating this year the 200th anniversary of
Sir William Rowan Hamilton \cite{Hami}, particularly here in Ireland.
One of his achievements was
the discovery of quaternions in 1843: he had thought for a long time
about possibilities to extend the representation of complex numbers
by two real components to a system with three real components, until he 
noticed that he had to proceed to four components to arrive at a sensible
system --- the first {\em non-commutative} (NC) algebra that was studied.

In the 20th century, the concept of non-commutativity
for space coordinates and momenta became standard in quantum physics,
but also the idea of NC space coordinates is about 60 years old \cite{Sny}. 
It experienced a powerful renaissance in 1998, triggered by
the identification of open strings at low energy with NC field
theory \cite{SeiWit}. NC spaces are also considered a promising approach to
quantum gravity \cite{DFR}. 
On the phenomenological side, it might for instance explain
the observation of high energy photons from far away galaxies,
beyond the energy threshold apparently predicted by the Standard Model
\cite{cosmo}. Here we study NC gauge theory in its own right.

In that framework, quantum mechanical position operators obey a 
commutation relation
of the form \ $ [\hat x_{\mu}, \hat x_{\nu} ] = i \Theta_{\mu \nu}$ ,
where we assume the non-commutativity tensor $\Theta$ to be
constant in (Euclidean) space-time. More precisely, we consider the case
of two commutative directions (which include the Euclidean time),
and an NC plane with the relation
$ [\hat x_{i}, \hat x_{j} ] = i \theta \epsilon_{ij}$ ~
($i,j \in \{ 1,2 \}$).

Field theory on such a space can be written in terms of our
usual (commutative) coordinates $x$, if all field multiplications are
performed by {\em star products,}
\begin{equation}
\phi (x) \star \psi (x) := \phi (x) \ \exp \Big( \, 
\frac{1}{2} \backderi \,\! _{i} \, \theta \, \epsilon_{ij} \,
\forderi \,\! _{j} \, \Big) \ \psi (x) \ .
\end{equation}
In particular, the action of pure $U(1)$ gauge theory takes the form
\begin{equation}
S[A] = \frac{1}{4 g^{2}} \int d^{4}x \ 
{\rm Tr} \, [ F_{\mu \nu} \star F_{\mu \nu} ] \ , \quad
F_{\mu \nu} = \partial_{\mu} A_{\nu} - \partial_{\nu} A_{\mu}
+ i [ A_{\mu} \star A_{\nu} -  A_{\nu} \star A_{\mu}] \ ,
\end{equation}
which is star-gauge invariant.
The Yang-Mills type self-interaction term is expected to yield
a ``photo-ball'' spectrum \cite{photoball} 
\footnote{The term ``photo-ball'' is an obvious analogue
to the glueball; to our knowledge, it first occurred in the work
by Fatollahi and Jafari quoted in Ref.\ \cite{photoball}.}.
It may modify the
photon dispersion relation at low energy, as it was observed
non-perturbatively in the NC $\lambda \phi^{4}$ model \cite{phi4},
as a consequence of UV/IR mixing effects \cite{UVIR}.
These effects drastically complicate the perturbative treatment. 
However, the $\beta$ function could be computed, 
suggesting asymptotic freedom \cite{afree}. 
There are a number of further perturbative
\cite{pertu} and semi-classical \cite{semicla} studies.

A formulation on a (fuzzy) lattice is possible \cite{RJS}, and it relates
the spacing $a$ on a $N \times N$ lattice to the NC parameter as 
\begin{equation}  \label{thetavsa}
\theta = \frac{1}{\pi}Na^{2} \ .
\end{equation}
Then the {\em Double Scaling Limit} (DSL), which takes simultaneously
\ $N \to \infty$ \ and \ $a\to 0$ \ at \
$\theta = const.$~, leads to a continuous NC plane of infinite extent.

But such a formulation is not immediately applicable
for simulations, in particular because of the request for
star-unitary link variables. 
However, there is an exact map \cite{TEKmap} of the NC 
$N \times N$ lattice onto the dimensionally reduced 
Twisted Eguchi-Kawai model \cite{TEK} with the action
\begin{equation}
S[U] = - \beta N \sum_{i \neq j} {\cal Z}_{ij} \, {\rm Tr} \,
[ U_{i} U_{j} U_{i}^{\dagger} U_{j}^{\dagger} ] \ , \quad
{\cal Z}_{12} = {\cal Z}_{21}^{*} = \exp \{ \pi i (N+1)/N \} \ ,
\quad N ~~ {\rm odd} \ ,
\end{equation}
where $U_{i} \in U(N)$, \ $i= 1,2$.
The (analogue of a) rectangular Wilson loop of sides $aI$ and $aJ$
is now given by
\beq
W_{ij} ( I \times J) = \frac{1}{N} {\cal Z}_{ij}^{IJ} \
{\rm Tr} \, \Big( U_{i}^{I} U_{j}^{J} U_{i}^{\dagger \, I} 
U_{j}^{\dagger \, J} \Big) \ .
\eeq
Mapping this quantity back to the lattice yields in fact
a sensible definition of a Wilson loop in the NC gauge theory \cite{IIKK}.
Note that NC Wilson loops are star-gauge invariant and
complex.


\section{Numerical results for the Double Scaling behaviour}\label{section2}


This mapping of the NC plane onto a matrix model (one in each lattice site
of the commutative plane) enables numerical simulations.
The next challenge is 
to identify the dimensional lattice spacing $a(\beta )$ in order to evaluate
observables in the DSL. The simple ansatz
\begin{equation} \label{dslansatz}
a \propto 1/\beta
\end{equation}
turned out to be successful, as we are going to
illustrate in a sequence of plots.
Along with relation (\ref{thetavsa}) it implies
$N/\beta^{2} = const.$
\footnote{In contrast, in $d=2$ we had to set $a \propto 1/ \beta^{2}$,
and therefore $N / \beta = const.$ \cite{NCQED2d}.}
We always deal with $N^{2} \times (N \pm 1)^{2}$ lattices, where
$N^{2}$ ( $(N \pm 1)^{2}$ ) is the lattice size in the NC 
(in the commutative) plane, and the NC plane is mapped onto a
twisted Eguchi-Kawai model.
We present results at $N=45, \, 55, \, 65,\, 71$ and $81$, and
the corresponding $\beta$ values are fixed such that
$N/\beta^{2} \equiv 20$ in all cases.
\footnote{We are also working on a systematic search
for Double Scaling by matching the data without any assumption about
the relation between $\beta$ and $a$. This represents a completely 
unbiased test of the DSL postulated here.}
This means that we are always in the weak coupling phase 
\cite{TEK,Lat04}.

\begin{figure}[htbp]
\includegraphics[angle=270,width=.48\linewidth]{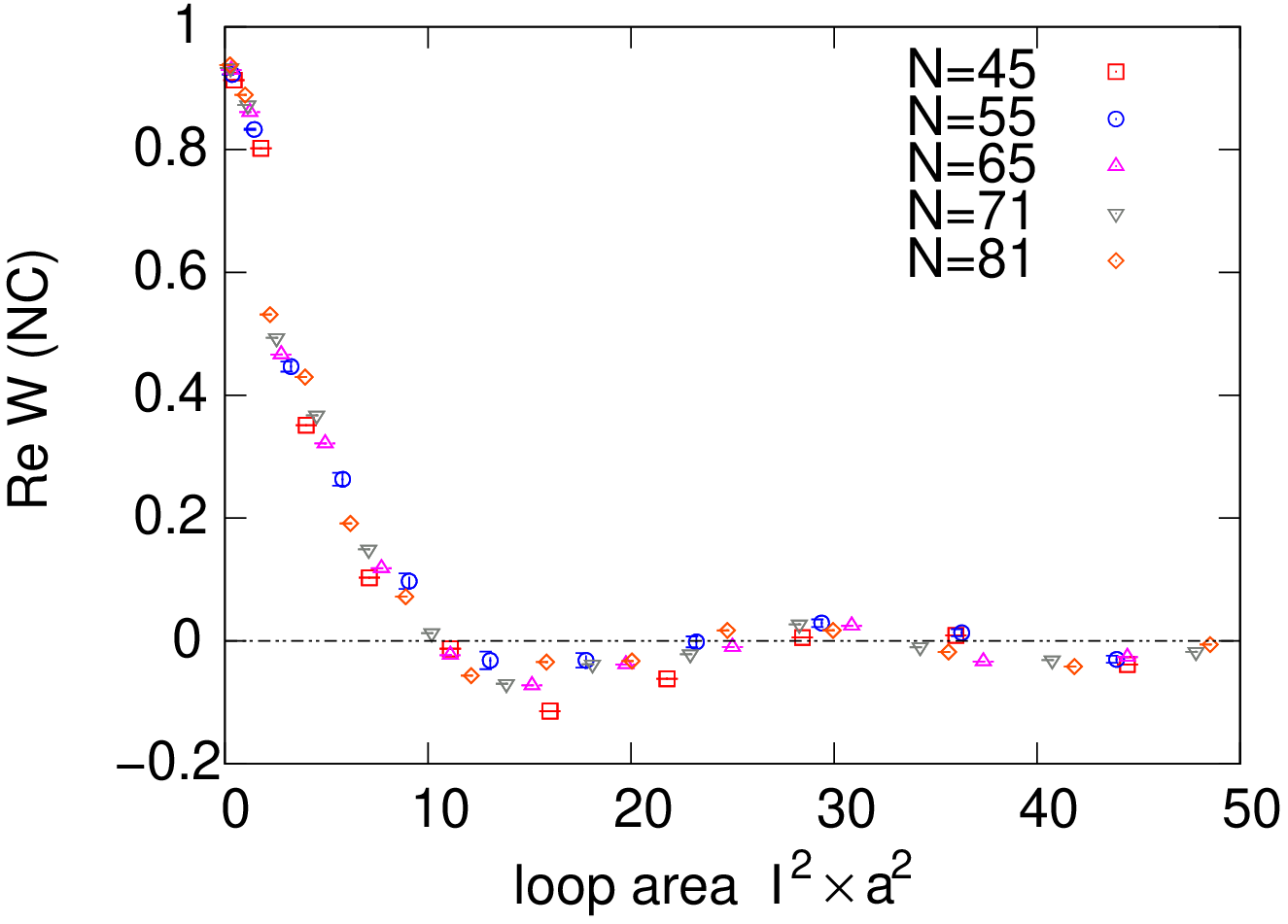}
\includegraphics[angle=270,width=.46\linewidth]{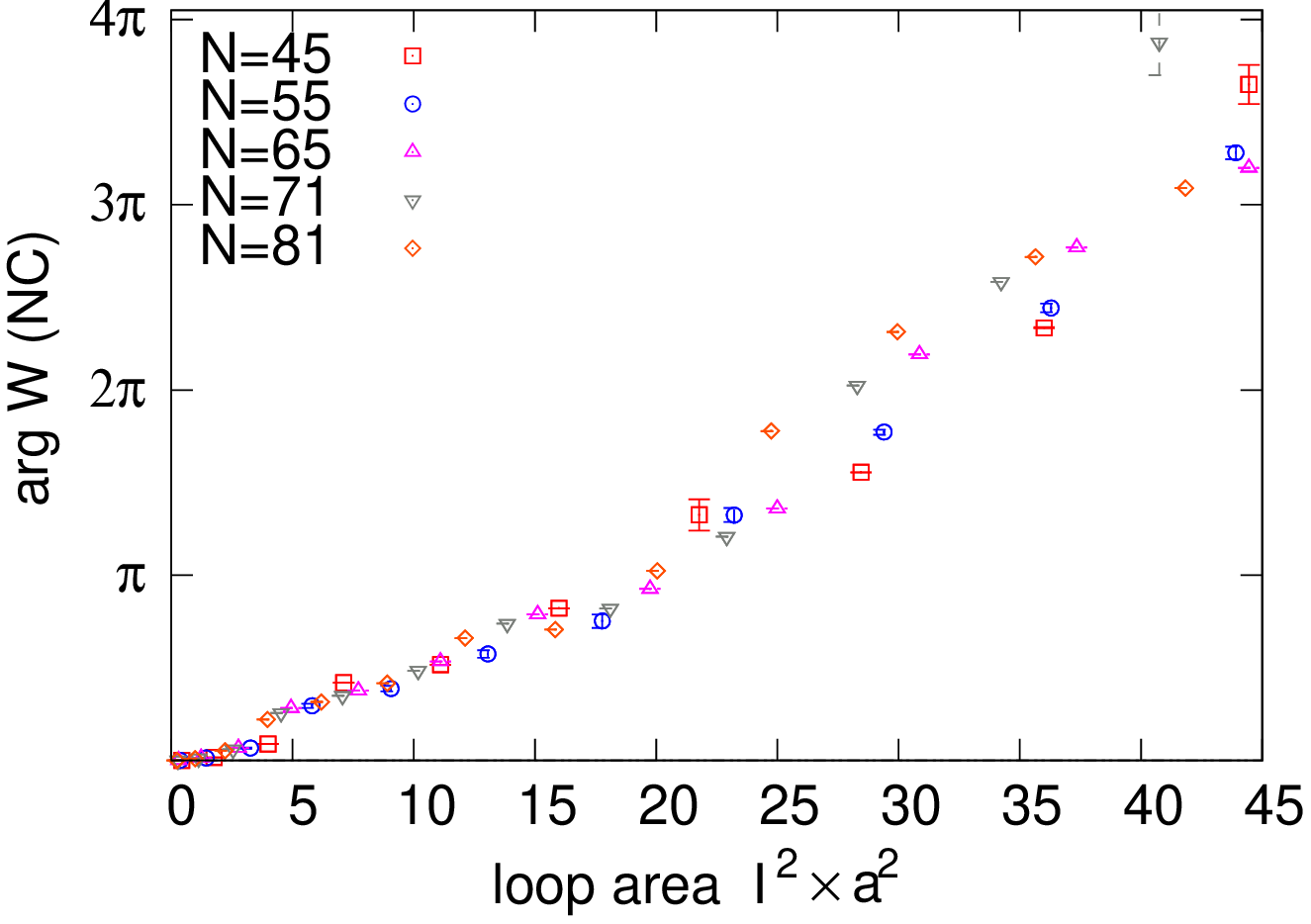}
 \caption{The Wilson loop in the NC plane: its real part (on the left)
decays at small area, and for larger areas it oscillates around zero.
In that regime, the complex phase (on the right) turns sizeable and 
it begins to grow in an (approximately) linear way.}
\label{W_NCfig}
\vspace*{-3mm}
\end{figure}

Figure \ref{W_NCfig} (on the left) shows the real part of the 
Wilson loop in the NC plane as a function of the loop area.
We see that the loops of the same area $I^{2} a^{2}$ reveal a convincing
Double Scaling, if we insert the ansatz (\ref{dslansatz})
(we chose the proportionality constant $ = 1$).
The same is true for the phase of these Wilson loops, as the plot on the
right-hand-side of Figure \ref{W_NCfig} shows.
Qualitatively this behaviour is similar to the Wilson loops in 2d NC
QED \cite{NCQED2d}: small loops are almost real and decay exponentially
as the area increases. On the other hand,
for large loops the real part oscillates around zero,
and the phase grows linearly. The latter property is reminiscent of
the Aharonov-Bohm effect, if one identifies $\theta $ with an inverse
magnetic field across the NC plane, as suggested by Peierls \cite{Pei}.

Next we consider correlation functions of Wilson loops, which are located
in the NC plane but separated in the (commutative) Euclidean time.
As examples, we show the correlation of the real parts of $4 \times 4$
loops (on the left), and of the phases of $2 \times 2$ loops
(on the right).
In both cases we still observe a decent Double Scaling. Its quality
improves as $N$ in increases, which indicates that our simple ansatz
(\ref{dslansatz}) does indeed describe the DSL limit asymptotically.

\begin{figure}
\hspace*{-3mm}
  \includegraphics[angle=270,width=.52\linewidth]{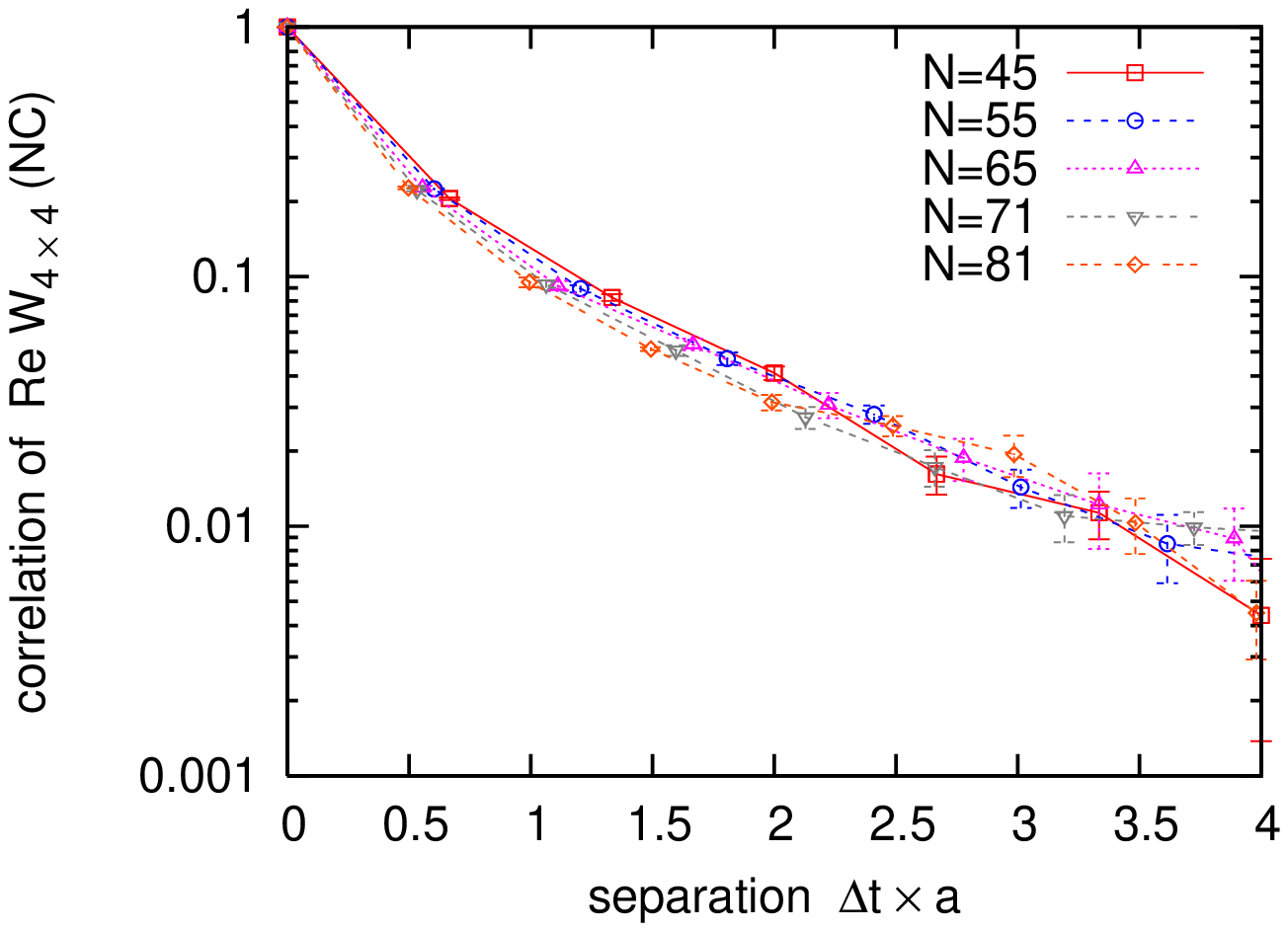}
  \includegraphics[angle=270,width=.52\linewidth]{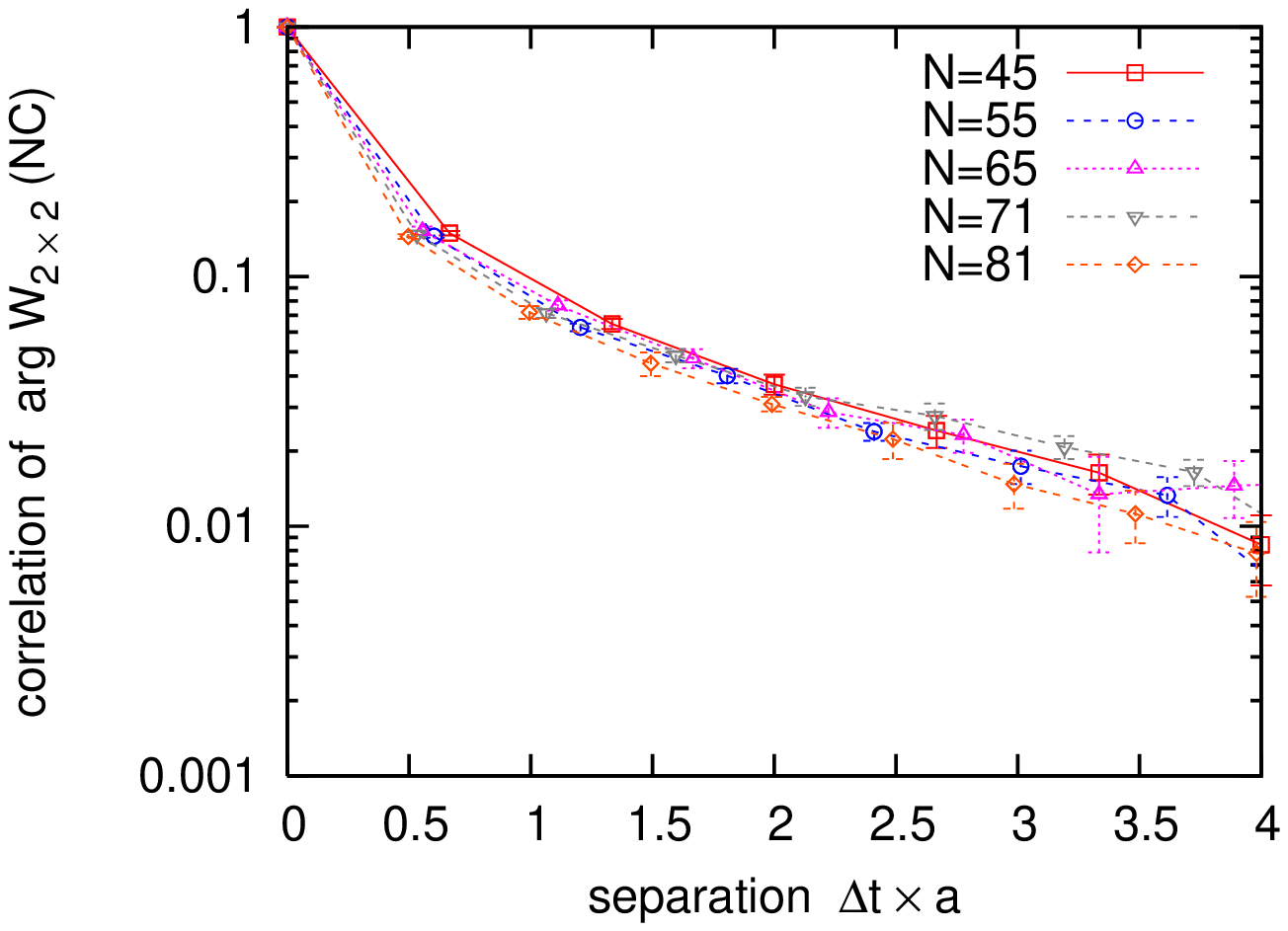}
 \caption{The correlation of Wilson loops in the NC plane, separated by a 
distance $a \Delta t$ in Euclidean time. Our examples are the
correlation the real parts of $4 \times 4$ Wilson loops (on the left), 
and of the phases of $2 \times 2$ loops (on the right).}
\label{W_NCcorfig}
\vspace*{-3mm}
\end{figure}

\begin{figure}
\hspace*{-3mm}
  \includegraphics[angle=270,width=.52\linewidth]{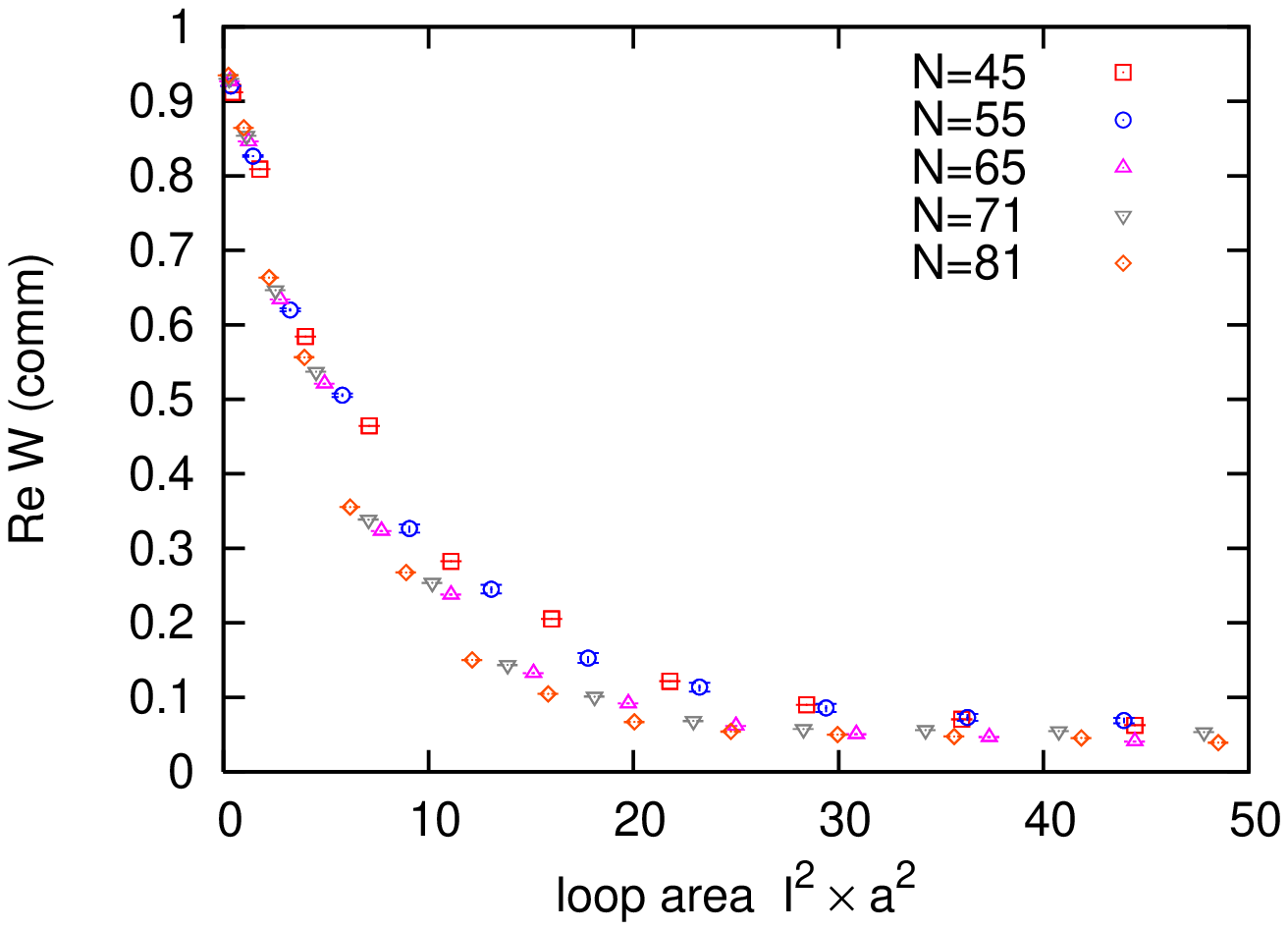}
  \includegraphics[angle=270,width=.52\linewidth]{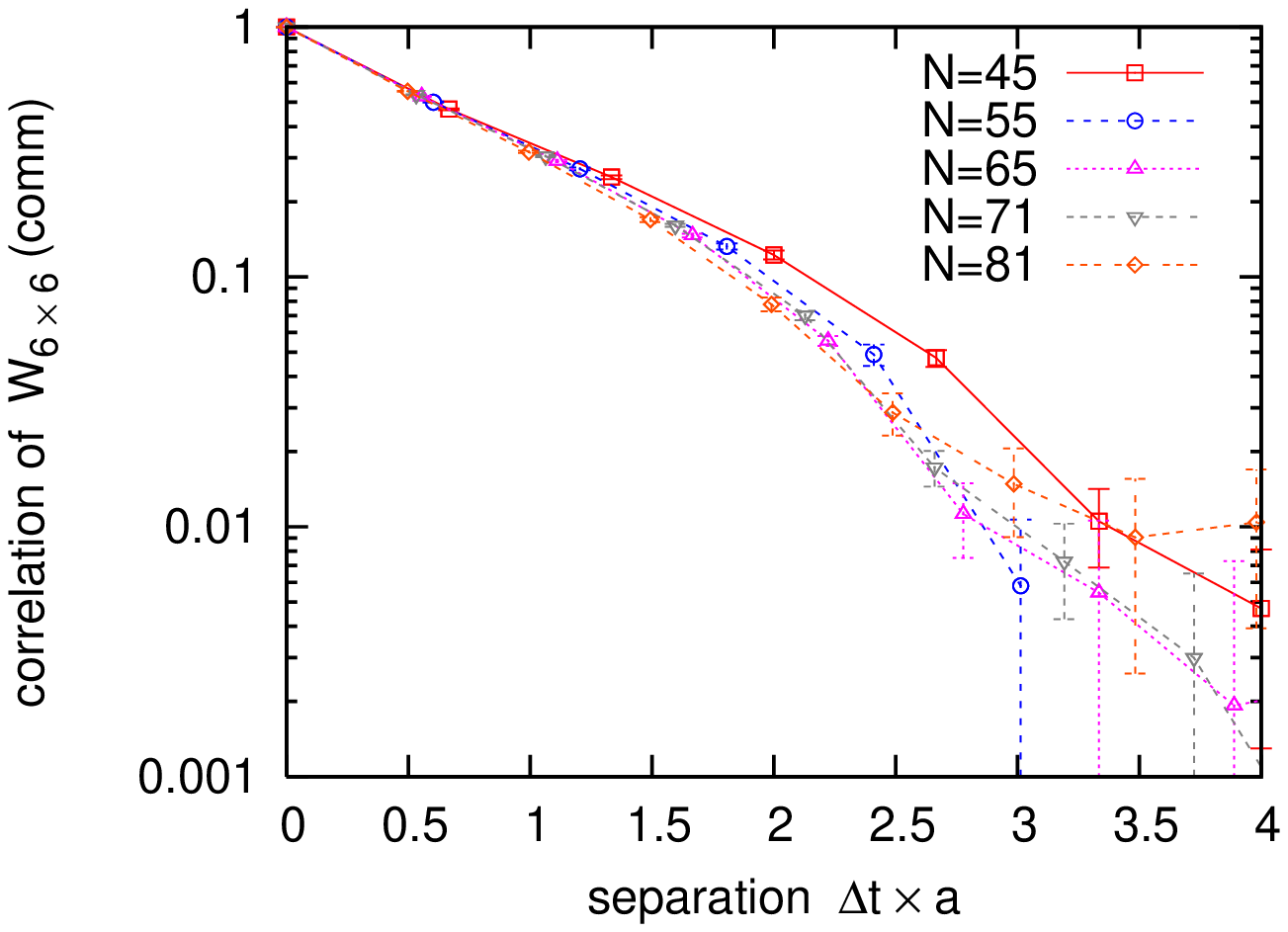}
 \caption{The Wilson loop in the commutative plane: the plot on the left depicts
the loops themselves (they are real) against the dimensional area. 
On the right we show the correlation of $6 \times 6$ loops, separated in
Euclidean time.}
\label{W_Cfig}
\vspace*{-3mm}
\end{figure}

Finally we also consider Wilson loops in the commutative plane.
These loops, as well as those in the mixed planes, are real
due to the reflection symmetry on the commutative axes.
From Figure \ref{W_Cfig} (on the left) we see that for loops in
this plane the convergence towards the DSL is a little more laborious,
but for our largest $N$ values it sets in also here.
On the right-hand-side of this Figure we also consider the correlator
of $6 \times 6$ loops in the commutative plane, again separated in
the Euclidean time, which confirms the above statement with respect
to the DSL.

\section{Conclusions and outlook}\label{section3}


We reported on our progress in a numerical investigation of pure
$U(1)$ gauge theory in a NC space-time. In particular, we presented 
results which reveal the asymptotic rule for the Double Scaling,
which takes the system to the continuum and to the infinite
volume at the same time; the entanglement of these limits can be viewed
in the light of the notorious UV/IR mixing. We observed that the simple 
ansatz (\ref{dslansatz}) for the lattice spacing works in a satisfactory way.

Hence the bases is now provided to study the observables of
physical interest and extrapolate them to the DSL. In particular, the
Wilson loop correlators ---
examples are shown in Figure \ref{W_NCcorfig} --- provide the 
bases for the evaluation of the ``photo-ball spectrum''.
We also hope to obtain results for the NC distortion of the photon
dispersion relation, which could then be confronted with
experimental data. Several high precision experiments dealing
with cosmic rays are about to measure the photon dispersion to
a very high accuracy, over a broad range of energies
(see for instance Ref.\ \cite{GLAST}).

While our results suggest the existence of a finite 
continuum theory, we note that perturbative calculations revealed
an infrared instability of the trivial vacuum \cite{IRinstab} (as long as 
the model is not rendered supersymmetric).
Indeed, we do observe numerically that the open Polyakov lines 
(which are star-gauge invariant as well)
acquire non-zero expectation values, in accordance with
perturbation theory. We therefore consider 
that we are actually probing a stable vacuum, which might be obtained
after the condensation of ``tachyons'' in the trivial vacuum. 
This issue shall be discussed in a forth-coming paper. \\


\vspace*{-3mm}

{\small
\noindent
{\bf Acknowledgements} \ \ 
Frank Hofheinz has contributed to this work in an early stage,
and Hinnerk St\"{u}ben has given us helpful advice regarding 
the parallelisation of our code.
We also thank Antonio Bigarini, Chong-Sun Chu,
Harald Grosse, Esperanza Lopez, Carmelo Perez Martin, Stam Nicolis, 
Denjoe O'Connor, Andrzej Sitarz, Richard Szabo and Alessandro Torielli
for stimulating discussions.
We are grateful for support by the ``Deutsche Forschungsgemeinschaft'' 
(DFG). Most computations were performed on the IBM p690 
clusters of the ``Norddeutscher Verbund f\"ur Hoch- und 
H\"ochstleistungsrechnen'' (HLRN).
}


\end{document}